\documentclass[%
 reprint,
superscriptaddress,
nofootinbib,
 amsmath,amssymb,
 aps,
]{revtex4-2}

\usepackage{graphicx}%
\usepackage{dcolumn}%
\usepackage{bm}%
\usepackage{subcaption}
\usepackage{siunitx}

\usepackage{physics}
\begin{document}

\preprint{APS/123-QED}

\title{Standalone gradient measurement of matrix norm\\for programmable unitary converters}

\author{Yoshitaka Taguchi}
\email{ytaguchi@ginjo.t.u-tokyo.ac.jp}
\affiliation{Research Center for Advanced Science and Technology, The University of Tokyo, Tokyo, Japan}

\author{Yasuyuki Ozeki}
\affiliation{Research Center for Advanced Science and Technology, The University of Tokyo, Tokyo, Japan}

\date{\today}%

\begin{abstract}
Programmable unitary converters are powerful tools for realizing unitary transformations, advancing fields of computing and communication. The accuracy of these unitary transformations is crucial for maintaining high fidelity in such applications. However, various physical artifacts can impair the accuracy of the synthesized transformations.
A commonly employed approach uses the system's gradient to restore accuracy. Matrix norm is used to define error between matrices, and minimization of this norm using the gradient restores the accuracy.
Although this gradient can indeed be physically measured using external equipment, it leads to a rather complex optical system.
In this study, we propose a standalone method for measuring matrix norm gradients, where `standalone' means that no additional optical equipment is needed.
This method is based on the mathematical fact that the central difference, which is generally used for the approximation of differentiation, can yield exact differentiation for any unitary converters.
Furthermore, we introduce a new matrix distance that is suitable for optimizing unitary converters which use intensity detectors at the output.
This distance also yields the exact differentiation with the central difference.
Numerical analysis demonstrates that our method exhibits orders of magnitude higher tolerance to measurement noise than prior similar approaches.
\end{abstract}

\maketitle

\section{Introduction}
\label{sec:intro}
Programmable unitary converter is advancing photonic quantum information processing \cite{Carolan2015,Ding2017,Elshaari2020,Pelucchi2022,Dong2023}, with its applications extending to machine learning \cite{Shen2017,Zhang2021,Ashtiani2022,Pengfei2022}, communications \cite{Fontaine2012,Annoni2017,Zhou2020,Romero2023,SeyedinNavadeh2023}, and blockchain \cite{Pai2023}. Numerous approaches have been proposed for the realization of a programmable unitary converter on optical platforms. A significant challenge in this field is handling imperfections in optical components, such as beam splitters and mode multiplexers. These imperfections can cause errors in the synthesized unitary transformation, leading to lower system fidelity. Unitary transformations are synthesized with passive unitary components and variable phase shifters in the physical device, with the realized transformation being parameterized through the amount of phase shifts in these phase shifters.

Gradient-based calibration methods serve as potent approaches for system calibration and the realization of accurate unitary transformation \cite{Pai2019,Zhou2020,Alexiev2021,Taguchi2023}. The error between the target unitary matrix and the matrix realized by the actual physical unitary transformer is defined by a matrix norm. The gradient of this matrix norm is then used to minimize the error. Such error minimization methods have been explored for major programmable unitary converter architectures.
However, the gradient of the matrix norm cannot be directly obtained in physical devices. Though the gradient can indeed be physically obtained by propagating light in the reverse direction \cite{Hughes2018,Fontaine2019,Zhou2020PhotonRes,Pai2023Science}, or backward-propagating error can be directly acquired through another optical path \cite{Ohno2022}, these approaches increase the complexity of the optical system. Our previous study approximated the gradient using the forward finite difference method \cite{Taguchi2023}, while it requires a smaller parameter difference to obtain a more accurate transformation. Using such a small difference is susceptible to system noise, and this limits the achievable accuracy.

In this study, we propose a standalone gradient-measurement method for programmable unitary converters that operates without any additional equipment. This method is based on the central difference approximation of the gradient, and we mathematically show that it can produce a value exactly equivalent to the analytical gradient of the matrix norm for any type of programmable unitary converter.
This property is supported by the fact that linear phase sweep in the programmable unitary converter generates sinusoidal change in the matrix norm.
While deriving the gradient, this property also offers a standalone correction method for the linearity of the phase shifters.
We numerically demonstrate the synthesis of target unitary transformations for unitary converters based on few-redundant multi-plane light conversion (MPLC) architecture \cite{Taguchi2023,Taguchi2023CLEO} using the proposed gradient-measurement method.
We consider two major detection methods: coherent detection and intensity detection. For intensity detection, we propose a matrix distance that is designed to give an exact gradient when combined with the proposed gradient-measurement method.
The proposed distance has unimodal property, and we demonstrate a good configuration result with the few-redundant MPLC architecture.
We also show that our proposed method exhibits several orders of magnitude greater tolerance to noise compared to the previous finite difference-based method.
This work reduces the hardware requirements of unitary converters for gradient-based calibration methods while maintaining the system's simplicity.
We also expect that this work provides alternative methods to obtain gradients in training optical machine learning platforms \cite{Hughes2018,Zhou2020PhotonRes,Wright2022,Zheng2023,Pai2023Science}.

\section{Central difference method for matrix norm}
\label{sec:central_diff}
We introduce the central difference method for multivariate function and show that this method can yield the exact analytical gradient of the matrix norm for programmable unitary converters. 
The common central difference method approximates the gradient of a function using function values. The gradient approximation of a function $y(x)$ using a central difference is defined by $y'(x) \approx [y(x+h)-y(x-h)]/2h$. Typically, the error of this approximation is of order $O(h^2)$, and the approximation does not match the exact gradient. By applying this equation to every variable of a multivariate function, we can extend this method to a multivariate function.
Let $f(x_1, x_2, \cdots, x_n)$ denote a multivariate function. The gradient obtained by applying the central difference method to this function, $\nabla_\mathrm{c} f$, is expressed as follows:
\begin{equation}
\label{eq:def_central_diff}
\nabla_\mathrm{c} f =
\begin{bmatrix}
\frac{f(x_1 + h, x_2, \cdots, x_n) - f(x_1 - h, x_2, \cdots, x_n)}{2h} \\
\frac{f(x_1, x_2 + h, \cdots, x_n) - f(x_1, x_2 - h, \cdots, x_n)}{2h} \\
\vdots \\
\frac{f(x_1, x_2, \cdots, x_n + h) - f(x_1, x_2, \cdots, x_n - h)}{2h}
\end{bmatrix},
\end{equation}
where $h$ denotes the finite difference. 
For the matrix norm minimization in programmable unitary converters, the variables $(x_1, x_2, \cdots, x_n)$ correspond to the phase shifts in the phase shifters of the device, while the function $f(x_1, x_2, \cdots, x_n)$ corresponds to the square of the matrix norm $\norm{U-X(x_1, \cdots, x_n)}^2_F$.
Here, $U$ is the target unitary matrix we aim to realize, $X(x_1, x_2, \cdots, x_n)$ is the actual unitary matrix realized by the physical converter, and $\norm{\cdot}_F$ denotes the Frobenius norm, which is a commonly used matrix norm for discussing the unitary matrix optimization. The vector, $\nabla_\mathrm{c} f$, can be obtained physically by manipulating the phase shifters. For systems using coherent detection at the output, our objective is to minimize the matrix error $f = \norm{U-X}_F^2$ to ensure that the realized unitary matrix $X$ aligns as closely as possible with the target unitary matrix $U$. The gradient of the matrix norm is necessary for using gradient-based calibration methods.

\begin{figure}[t]
\centerline{\includegraphics[width=82mm]{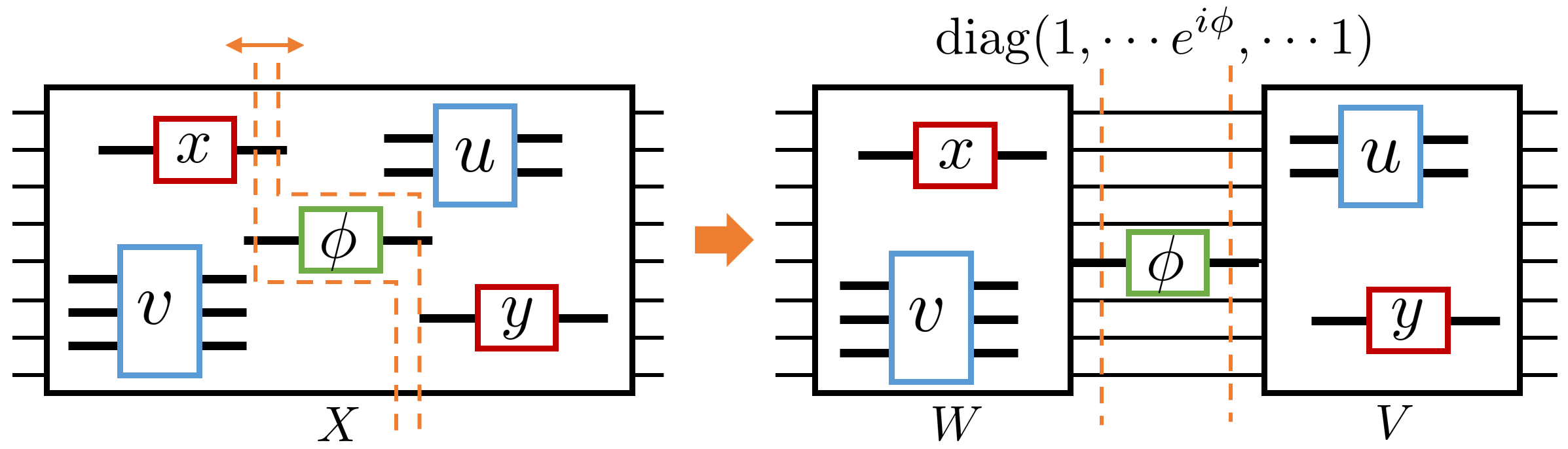}}
\caption{Schematic proof of the equation $X=V\mathrm{diag}(1, \cdots e^{i\phi}, \cdots 1)W$. We consider the situation where only one phase shifter is variable in the unitary converter device. 
Components $u, v$ represent passive unitaries, while components $x, y$ represent phase shifters with their phases fixed.
Of these, only the phase shifter $\phi$ is variable.
}
\label{fig:X_decomp_proof}
\end{figure}

To simplify Eq. (\ref{eq:def_central_diff}), we rewrite the finite difference term $f(x_1, \cdots, x_i + h, \cdots, x_n) - f(x_1, \cdots, x_i - h, \cdots, x_n)$ using a convenient universal property of $f(x_1, \cdots, x_n)$ that is derived from the unitary converter. We define $g(\phi)$, by fixing all variables in the multivariate function $f$ except one, thereby treating it as a single-variable function: $g(\phi) = f(x_1, \cdots, x_i = \phi, \cdots, x_n) = \norm{U-X(x_1, \cdots, x_i = \phi, \cdots, x_n)}^2_F$.
A unitary converter comprises passive linear optical components and phase shifters. Given this, each element of the matrix $X(\phi)$ can be written as 
\begin{equation}
\label{eq:X_elem}
X_{kl}(\phi) = a_{kl} e^{i\phi} + b_{kl},
\end{equation}
where $X_{kl}$ is the $(k, l)$-th element of the matrix $X$, $a_{kl}$ and $b_{kl}$ are complex constants. This relationship holds because there exist unitary matrices $V$ and $W$ such that $X(\phi) = V \mathrm{diag}(1, \cdots e^{i\phi}, \cdots, 1) W$, as demonstrated schematically in Figure \ref{fig:X_decomp_proof}.
Using Eq. (\ref{eq:X_elem}) and the synthesis of trigonometric functions, $g(\phi)$ can be simplified as follows:
\begin{equation}
\label{eq:single_func}
\begin{aligned}
g(\phi) &= \norm{U-X(\phi)}^2_F = \sum_{kl} \abs{U_{kl} - X_{kl}(\phi)}^2\\
&=\sum_{kl}\abs{U_{kl}-b_{kl}}^2+\abs{a_{kl}}^2-2\Re\qty[\overline{(U_{kl} - b_{kl})}a_{kl}e^{i\phi}]\\
&= A \sin(\phi + \alpha) + B,
\end{aligned}
\end{equation}
where $A, B, \alpha$ are real constants.

We derive the exact analytical gradient of the function $f=\norm{U-X}^2_F$ using the central difference. Using the central difference of $g(\phi)$ and after a few lines of trigonometric calculation, we obtain
\begin{equation}
\label{eq:cent_diff_sinc}
\frac{g(\phi+h)-g(\phi-h)}{2h} = \dv{g}{\phi} \cdot \mathrm{sinc}(h).
\end{equation}
Since this relation is valid for each element in the central difference vector as presented in Eq. (\ref{eq:def_central_diff}), it leads to the conclusion
\begin{equation}
\label{eq:def_grad_cent}
\nabla f = \frac{1}{\mathrm{sinc}(h)} \nabla_\mathrm{c} f,
\end{equation}
which means that the exact analytical gradient $\nabla f$ can be obtained from the physically measurable $\nabla_\mathrm{c} f$.
As this discussion is solely based on the fact that the unitary converter comprises passive linear optical components and phase shifters, it remains valid for any unitary converter architecture.
In addition, this discussion remains valid as long as the passive elements in Figure \ref{fig:X_decomp_proof} are linear. This implies that the loss in passive elements does not affect Eq. (\ref{eq:def_grad_cent}). Not only photonics system based on unitary converters, but also wide range of linear converters driven by phase shifters \cite{Zhu2022} can benefit from this method.
To minimize the effect of measurement noise, $h=\pi/2$ is optimal since noise in $f$ is divided by $\mathrm{sinc}(h)\cdot 2h$ in Eq. (\ref{eq:def_grad_cent}), while the gradient remains constant irrespective of $h$. Therefore, maximizing this term minimizes the amplification of noise by the division.

Although this method assumes the linearity of the phase shifters, such linearity can be readily ensured through pre-calibration, as suggested by equation Eq. (\ref{eq:single_func}).
In other words, the signal to the phase shifters should be adjusted so that the norm value varies exactly in a sine-function manner during a linear sweep of the signal to the phase shifter.
It is worth noting that a similar formula cannot be obtained by other finite differences, namely, the forward difference $[g(\phi + h) - g(\phi)]/h$ and the backward difference $[g(\phi) - g(\phi - h)]/h$, highlighting the optimality of the central difference for this problem.

\section{Phase-insensitive distance tailored to central difference method}
In this section, we consider a case where intensity detectors are used at the output of the device and introduce a new distance tailored to our method.
For the application of the method discussed in Section \ref{sec:central_diff}, coherent detectors are needed because the calculation of matrix error, defined by the Frobenius norm, requires the measurement of complex amplitude at the output.
However, some applications, such as quantum optics using photon counters, rely solely on intensity information from the output.
Evaluating matrix error by the Frobenius norm is unsuitable, as its value varies with changes in the output phases.
To address this issue, our previous study proposed a matrix distance using a variant of the Frobenius norm, which remains invariant under phase shifts in the output \cite{Taguchi2023,Taguchi2023APLS}. The distance is expressed as:
\begin{equation}
\label{eq:prev_phase_insensitive}
d(U, X) = \sum_{ij} \qty(\delta_{ij} - \abs{\qty[XU^\dag ]_{ij}})^2.
\end{equation}
However, this definition lacks the convenient property that the value of the norm adopts the form of a sine function with an offset, as seen in Eq. (\ref{eq:single_func}). We can show that $\abs{\qty[X(\phi)U^\dag]_{ij}}^2 = A\sin(\phi + \alpha) + B$, where $A, B, \alpha$ are real constants, in a manner similar to the derivation of Eq. (\ref{eq:single_func}). Nevertheless, the definition of $d(U, X)$ includes a term $\abs{\qty[XU^\dag]_{ij}}$, and this term cannot be written using a pure sine function.

We propose an alternative phase-insensitive distance that can be expressed as a pure sine function with a constant offset. Although heuristic and top-down, we slightly modify Eq. (\ref{eq:prev_phase_insensitive}) to ensure that all $\phi$-dependent terms are expressed using sine functions.
The proposed distance is given by:
\begin{equation}
\label{eq:new_phase_insensitive}
d'(U, X) = \sum_{ij} \qty|\delta_{ij} - \abs{\qty[XU^\dag ]_{ij}}^2|.
\end{equation}
Since $0 \leq \abs{\qty[XU^\dag ]_{ij}}^2 \leq 1$, $d'(U, X)$ is non-negative and equals 0 if and only if $d(U, X) = 0$. In addition, this distance inherits the unimodality \cite{Taguchi2023} of original phase-insensitive distance $d(U, X)$. The proof of this unimodality is given in the Appendix. As evident from Eq. (\ref{eq:v_simple}) in the proof of unimodality, the range of the distance is obtained as $0 \leq d'(U, X) \leq 2N$. By defining $g(\phi)=d'(U, X(\phi))$, we can derive the same expression as in Eq. (\ref{eq:cent_diff_sinc}) for this distance, and the central difference method provides the exact analytical gradient.
This distance can also be used to pre-calibrate and ensure the linearity of the phase shifters for cases where intensity detectors are used at the output, as its value also varies sinusoidally.

\section{Numerical simulation}
\subsection{Setup}
\begin{figure}[hbtp]
\centerline{\includegraphics[width=75mm]{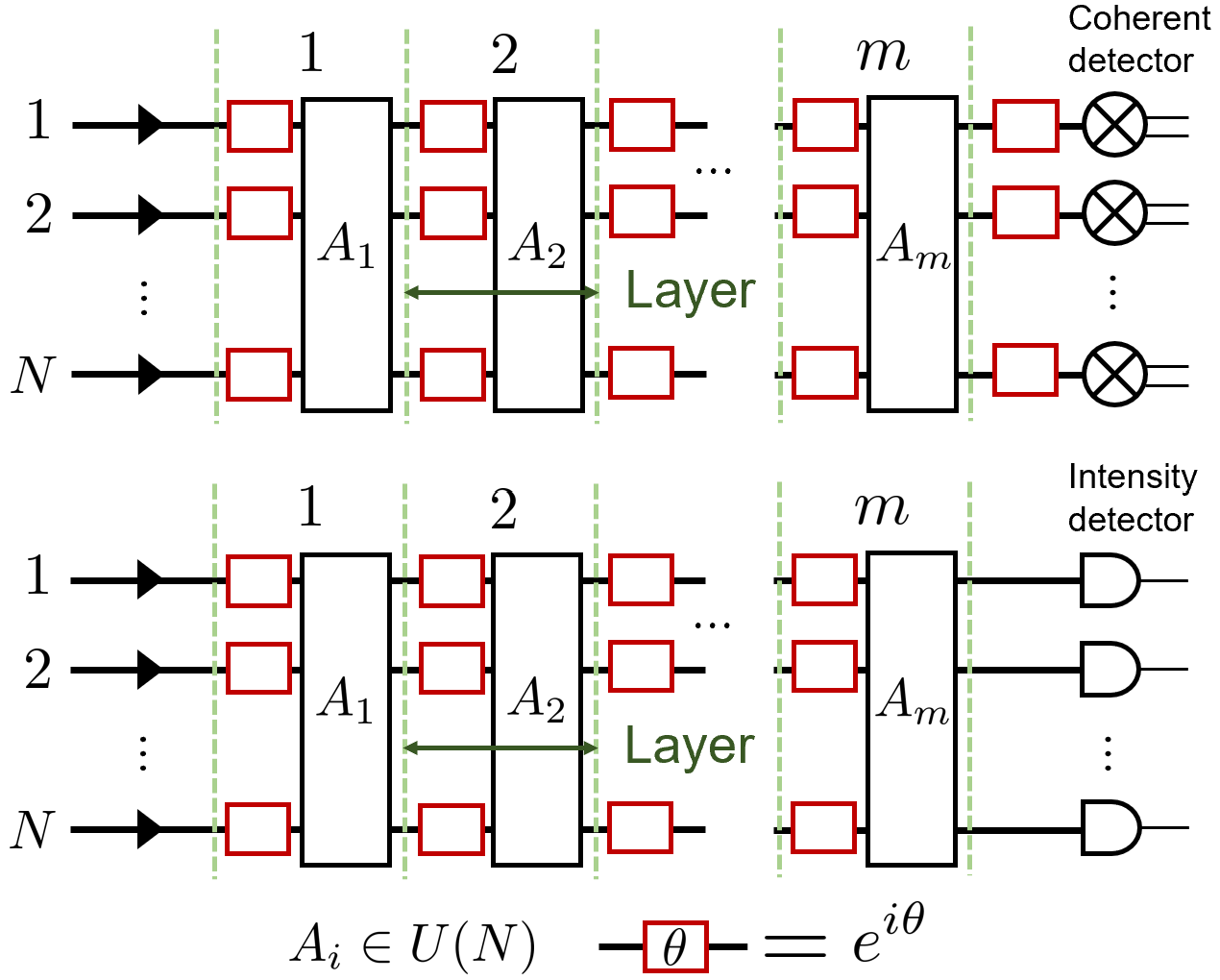}}
\caption{Schematic representation of the $N\times m$ MPLC architecture. It showcases coherent detectors and intensity detectors at the output. The number of layers is specified by $m$. Each layer consists of an $N$-port fixed unitary converter represented by $A_i$, followed by an array of $N$ single-mode phase shifters.}
\label{fig:MPLC}
\end{figure}

In our numerical simulations, we employ the few-layer redundant MPLC architecture \cite{Taguchi2023,Taguchi2023CLEO}.
This architecture has demonstrated its capability to be configured to a target unitary conversion using an iterative optimization method that leverages gradient information.
Figure \ref{fig:MPLC} illustrates this architecture with two different detectors at the outputs; one with coherent detector, and the other with intensity detector. The general MPLC architecture \cite{Morizur2010,Labroille2014} features $m$ layers for an $N$ port unitary transformation.
Each layer is comprised of an $N$-port fixed unitary converter $A_i$ and an array of $N$ single-mode phase shifters.
When using coherent detectors at the output, we append an additional phase shifter array following the final layer.
Given our usage of the few-layer redundant setup, the layer count $m$ is set to $N+1$. For our simulations, we set $N=8$.
Each distinct phase shift within the phase shifters is denoted by a real parameter variable, and all the phase shifts are collectively expressed as a vector $\vb{p}$.

To define the matrix error, the normalized cost function $\mathcal{L}$ \cite{Taguchi2023} between two matrices is defined as:
\begin{equation}
\label{eq:def_cost_func}
\mathcal{L}(\vb{p})= \left\{
\begin{aligned}
&\frac{1}{4N}\norm{U-X(\vb{p})}^2_F & (\text{Coherent detection}) \\
&\frac{1}{2N}d'(U, X(\vb{p})) & (\text{Intensity detection})
\end{aligned}
\right.
\end{equation}
so that $0 \leq \mathcal{L} \leq 1$ is always satisfied.
Here, $\vb{p}$ is a real parameter vector whose element corresponds to the amount of phase shift in the unitary converter,
$U$ represents the target matrix we aim to achieve and $X(\vb{p})$ denotes the unitary matrix physically produced by the parameter vector $\vb{p}$.
We employ numerical optimization to minimize $\mathcal{L}(\vb{p})$. At the start of the optimization, parameters are initialized using a uniform distribution ranging from $0$ to $2\pi$, and the target unitary matrix $U$ is sampled from the Haar measure using the \texttt{stats} module of SciPy \cite{ScipyNmeth}. In the MPLC architecture, the matrix $A_i$ for $1 \leq i \leq m$ is also sampled from Haar measure. After initializing the parameters and the matrix, the cost function $\mathcal{L}$ is optimized using the quasi-Newton optimization method, specifically, the limited-memory Broyden-Fletcher-Goldfarb-Shanno (L-BFGS) algorithm \cite{Flet1987} implemented in \texttt{optimize} module of SciPy \cite{ScipyNmeth}.
This method starts from the initial parameters and modifies them at each step until convergence to the local minimum, where $d\mathcal{L}/d\vb{p}=\vb{0}$. The optimization is run 64 times while changing the initial parameters to investigate the statistical behavior.
Matrices $U$ and $A_i$ are sampled at the each optimization. The gradient, which is essential for this optimization algorithm, is provided numerically either by central difference method or forward difference method for comparison.

\subsection{Results}
\begin{figure}[hbtp]
\centerline{\includegraphics[width=73mm]{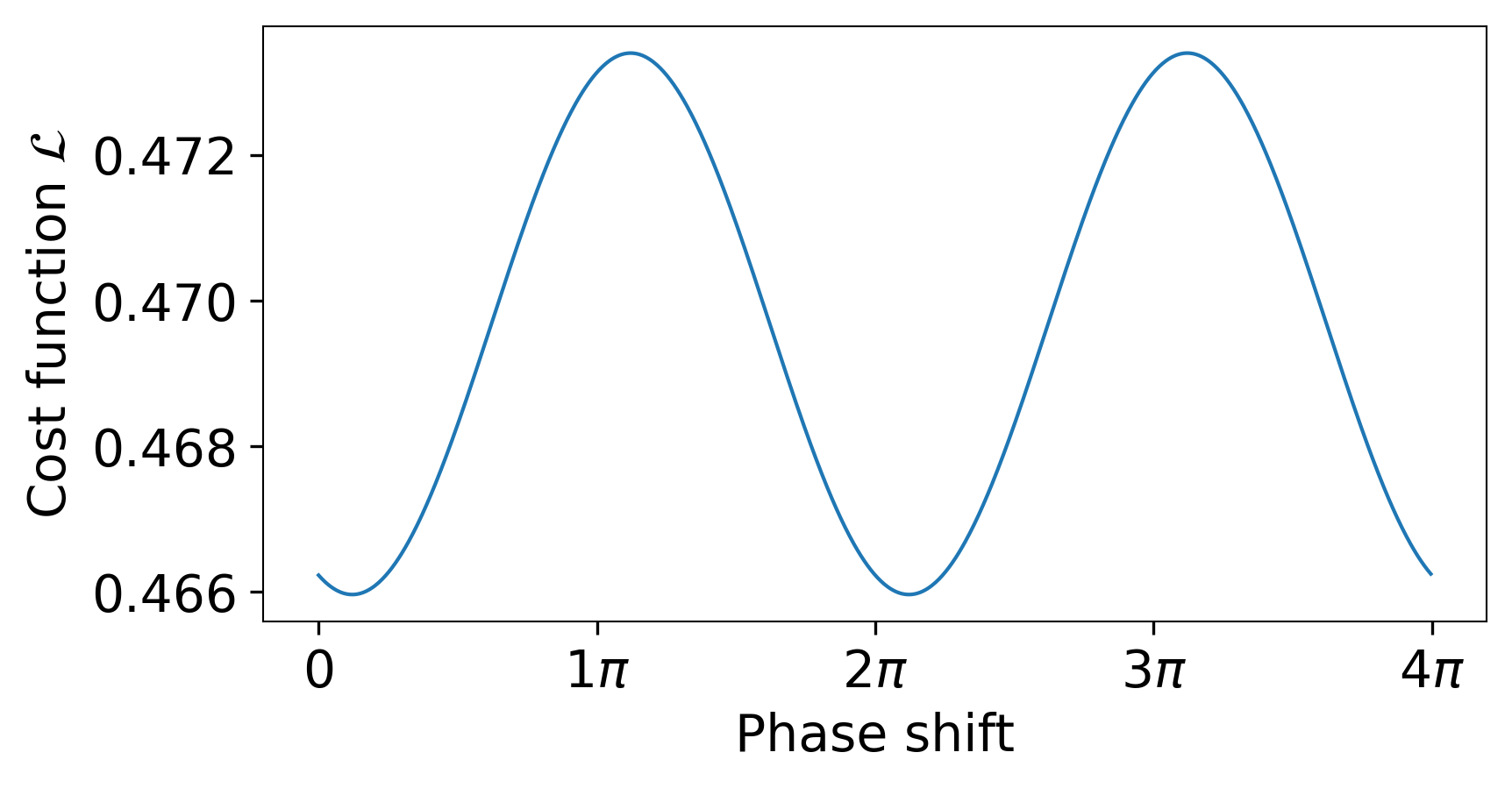}}
\caption{Variation of the cost function $\mathcal{L}(\vb{p})$ in response to changes in the phase shift of a single phase shifter, with all other phase shifters remaining fixed.}
\label{fig:phase_shift_sine}
\end{figure}

Figure \ref{fig:phase_shift_sine} illustrates the relationship between the cost function $\mathcal{L}(\vb{p})$ and the phase, when one phase shifter is varied while all other phase shifters are held fixed. As formulated in Eq. (\ref{eq:single_func}), the relation is expressed by a sine function with an offset.

\begin{figure}[hbtp]
\centerline{\includegraphics[width=82mm]{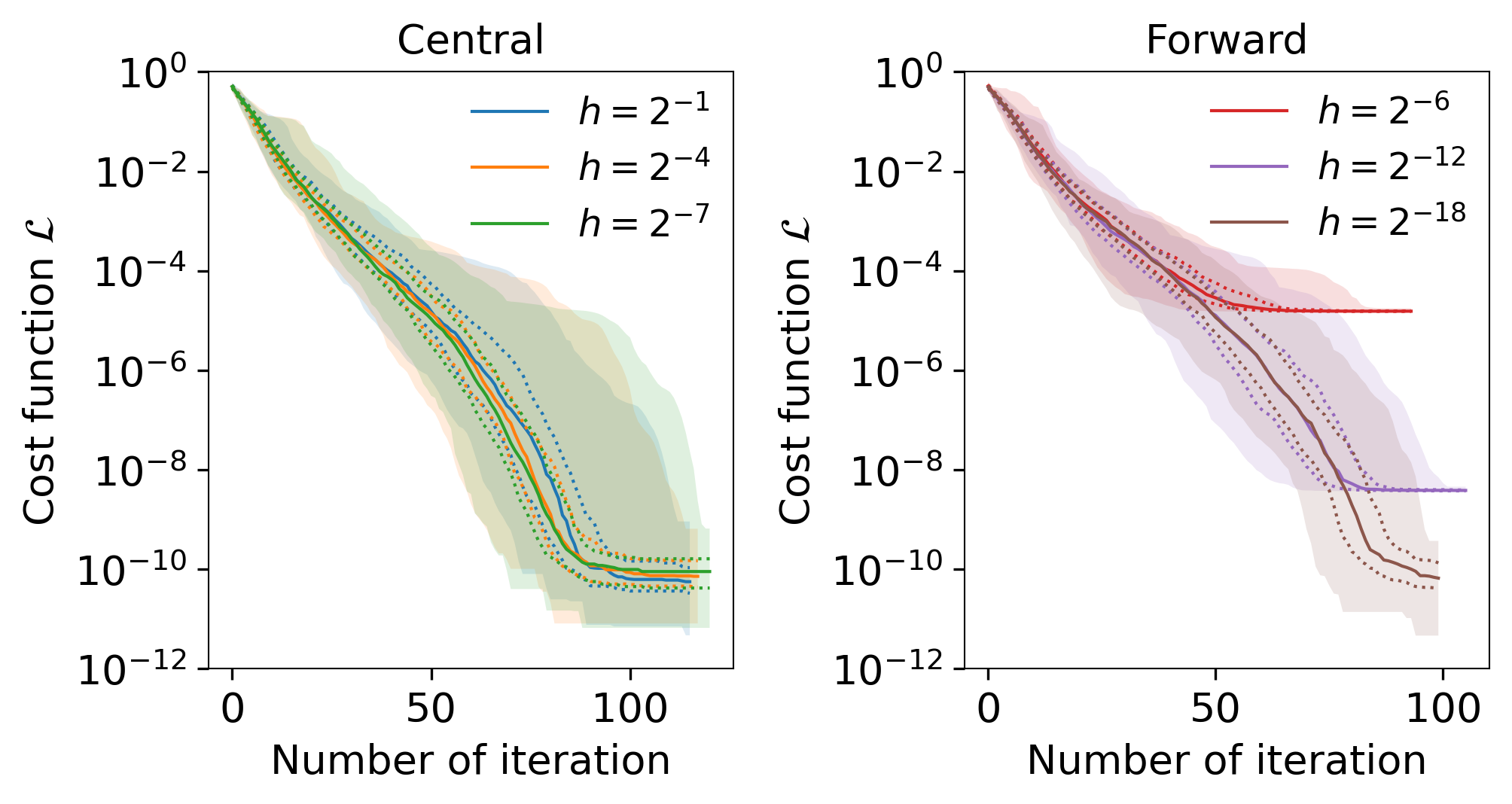}}
    \caption{Convergence comparison for cost function minimization in coherent detection between central and forward difference method. The vertical axis shows the the cost function $\mathcal{L}$ and the horizontal axis shows the number of iterations. The shaded area represents the minimum and maximum values, the solid line represents the median, and the dotted line represents the 25\% and 75\% quantiles over 64 optimization trials.}
\label{fig:forward_acc_comp}
\end{figure}

Figure \ref{fig:forward_acc_comp} illustrates convergence plots of the cost function minimization for coherent detection, comparing the use of central and forward difference methods for gradient computation. The results for the forward difference method were computed based on our previous report \cite{Taguchi2023}. We varied the finite difference $h$ to showcase its effect on gradient approximation error.
The convergence plot of the cost function is recorded for 64 optimization trials. The shaded area shows the range of minimum and maximum values, the dotted line shows the 25\% and 75\% quantiles, and the solid line shows the median of the trials.
Since the central difference method gives the exact gradient, irrespective of the value of $h$, the statistical behavior of the convergence remains constant across different $h$. In contrast, the convergence when using forward difference is significantly influenced by $h$, due to approximation errors it introduces.
To ensure results with accuracy limited by numerical precision, the forward finite difference method requires $h \leq 2^{-18}$.

\begin{figure}[htbp]
\centerline{\includegraphics[width=82mm]{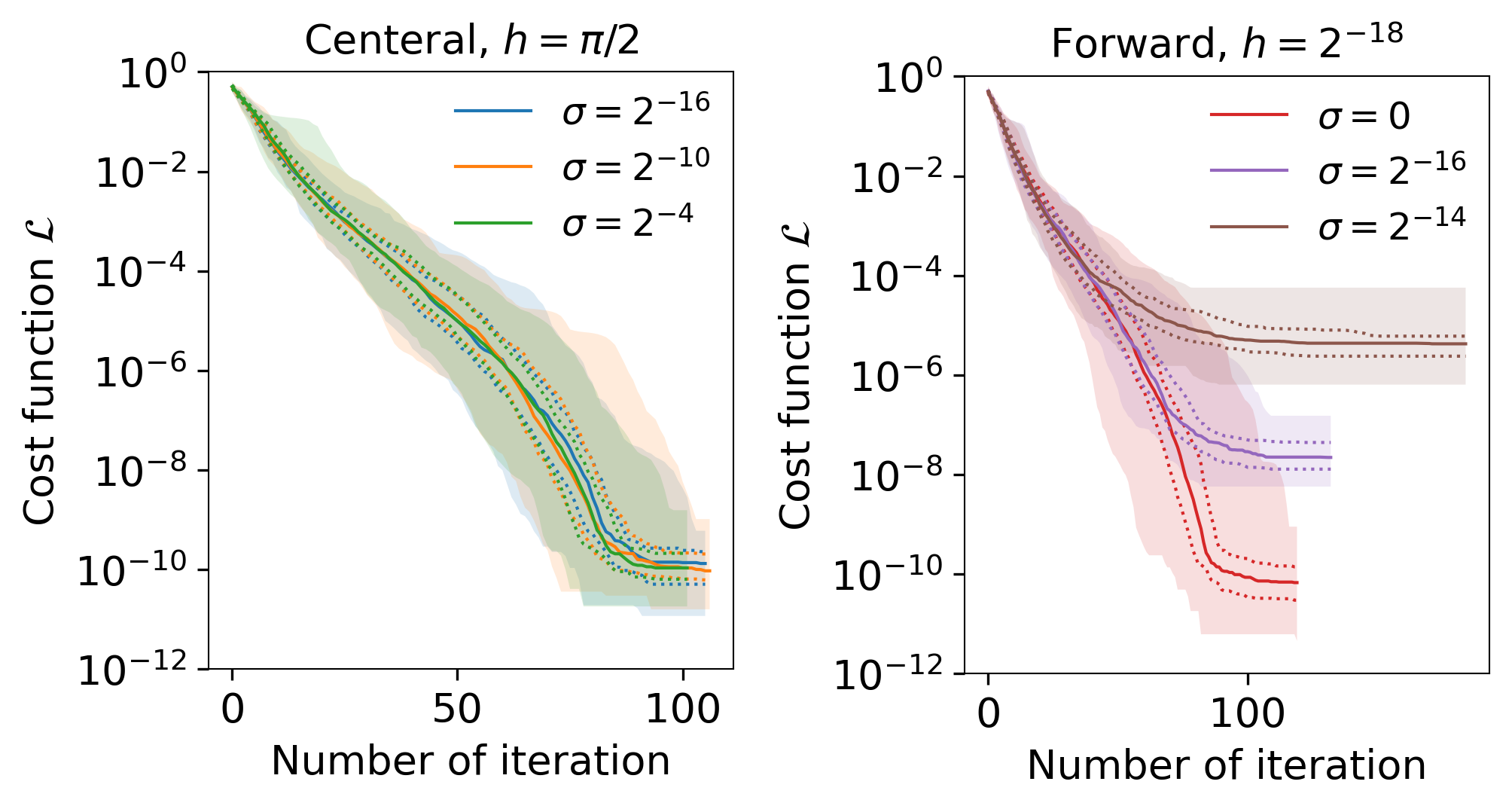}}
\caption{Comparison of convergence under noise using forward difference and central difference.}
\label{fig:noise_tolerance_comp}
\end{figure}

To contrast noise tolerance, we introduced noise into the measured matrix norm and conducted optimization. We replaced the term $f(x_1, \ldots, x_n)$ with $f(x_1, \ldots, x_n) + \varepsilon^2 = \norm{U-X}^2_F + \varepsilon^2$, where $\varepsilon \sim \mathcal{N}(0, \sigma^2)$ denotes noise following a normal distribution with mean $0$ and variance $\sigma^2$. 
Figure \ref{fig:noise_tolerance_comp} shows the convergence plots when the variance of added noise is varied. Optimization using forward difference with $h=2^{-18}$ is influenced by small noise as low as $\sigma=2^{-16}$, in comparison to the noiseless scenario where $\sigma=0$. The resulting accuracy is degraded by several orders of magnitude. This degradation can be attributed to the fact that any finite difference scheme involves division by the finite difference, and a small finite difference can cause significant noise amplification due to this division. Nevertheless, this problem can be mitigated by using a large finite difference $h$ with the proposed method.
As shown in Figure \ref{fig:noise_tolerance_comp}, optimization using the central difference with $h=\pi / 2$ is unaffected by noise. The scenario with large noise, $\sigma=2^{-4}$, exhibits similar statistical behavior to the scenario with small noise, $\sigma=2^{-16}$. This demonstrates a significantly higher level of noise tolerance compared to the use of forward difference.

\begin{figure}[htbp]
\centerline{\includegraphics[width=82mm]{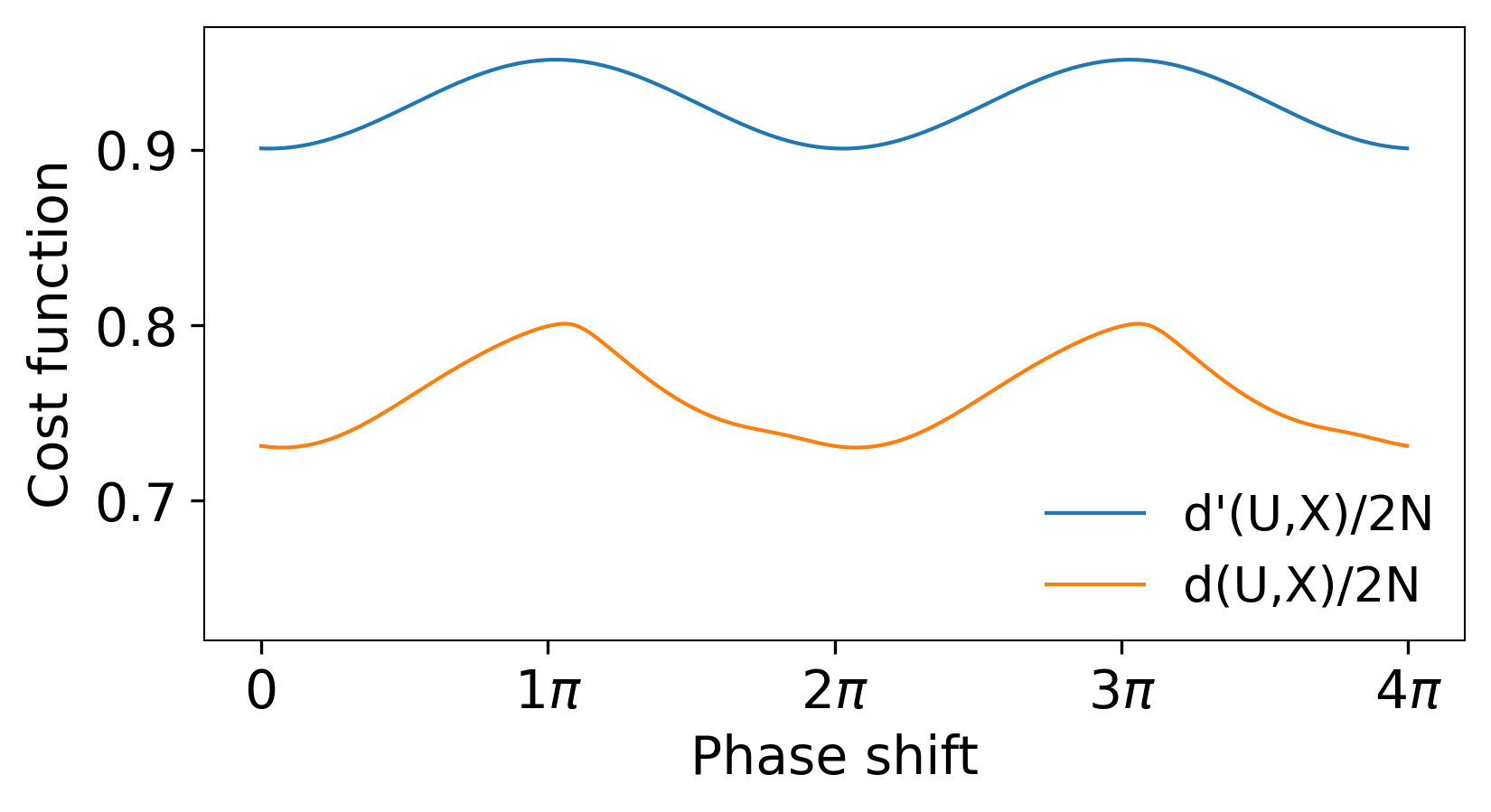}}
\caption{Comparison of the variation in the phase-insensitive distances $d(U, X)$ and $d'(U, X)$, normalized so their values range between $0$ and $1$.}
\label{fig:phase_insensitive_dist_comp}
\end{figure}

We compare the variation of the phase-insensitive distances $d(U, X)$ and $d'(U, X)$. Figure \ref{fig:phase_insensitive_dist_comp} illustrates the comparison of the normalized cost functions $d(U, X(\vb{p}))/2N$ and $d'(U, X(\vb{p}))/2N$. For this comparison, the phase of one phase shifter is varied while all other phase shifters remain fixed. As evident from the figure, the distance $d(U, X)$ does not exhibit a sine function form with an offset, indicating that the central difference method does not yield exact differentiation. In contrast, the proposed distance $d'(U, X)$ is expressed by a sine function with offset.

\begin{figure}[hbtp]
\centerline{\includegraphics[width=82mm]{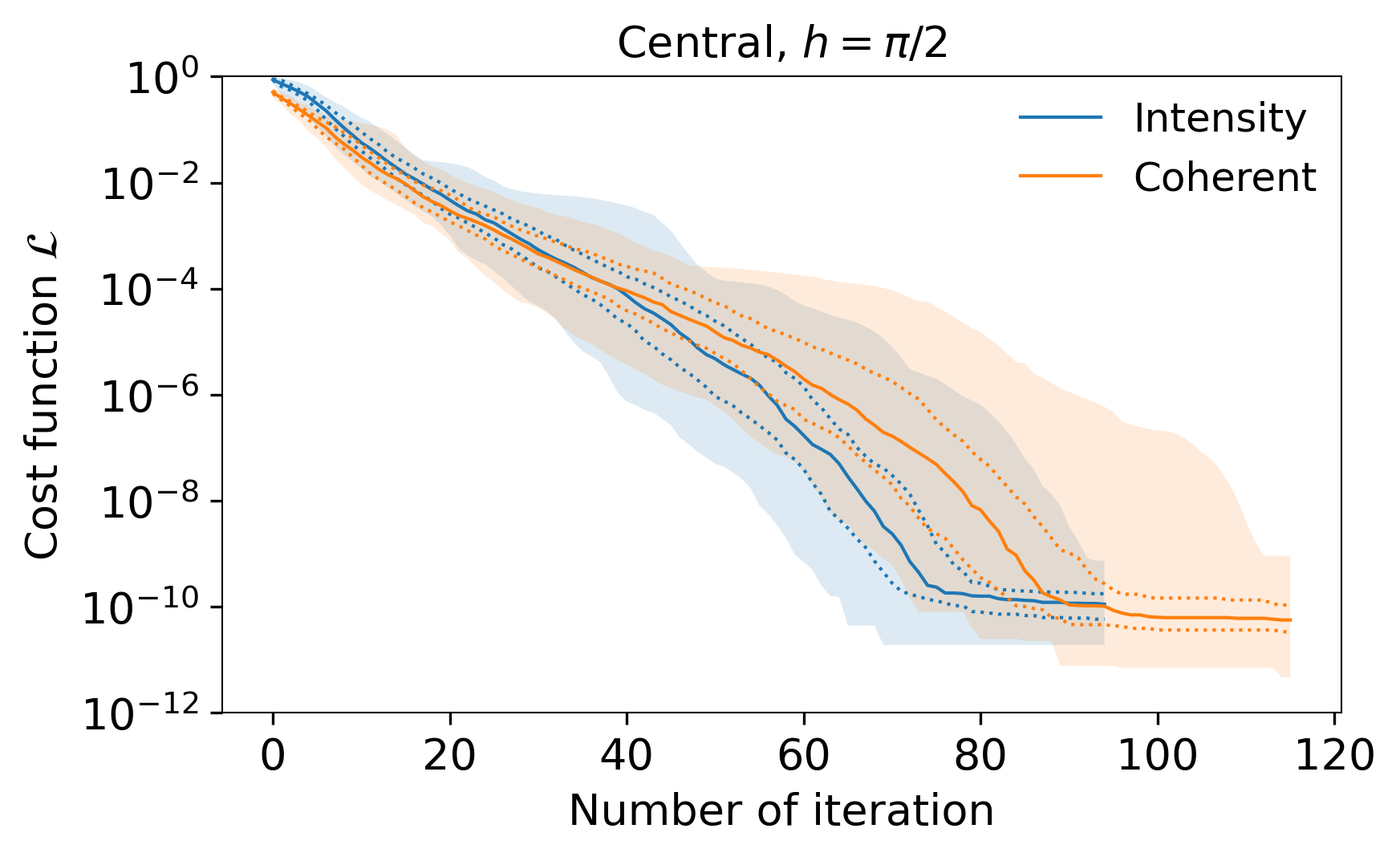}}
\caption{Comparison of convergence for cost function minimization in coherent detection and intensity detection.}
\label{fig:phase_insensitive_convergence_comp}
\end{figure}

\begin{figure}[hbtp]
\centerline{\includegraphics[width=75mm]{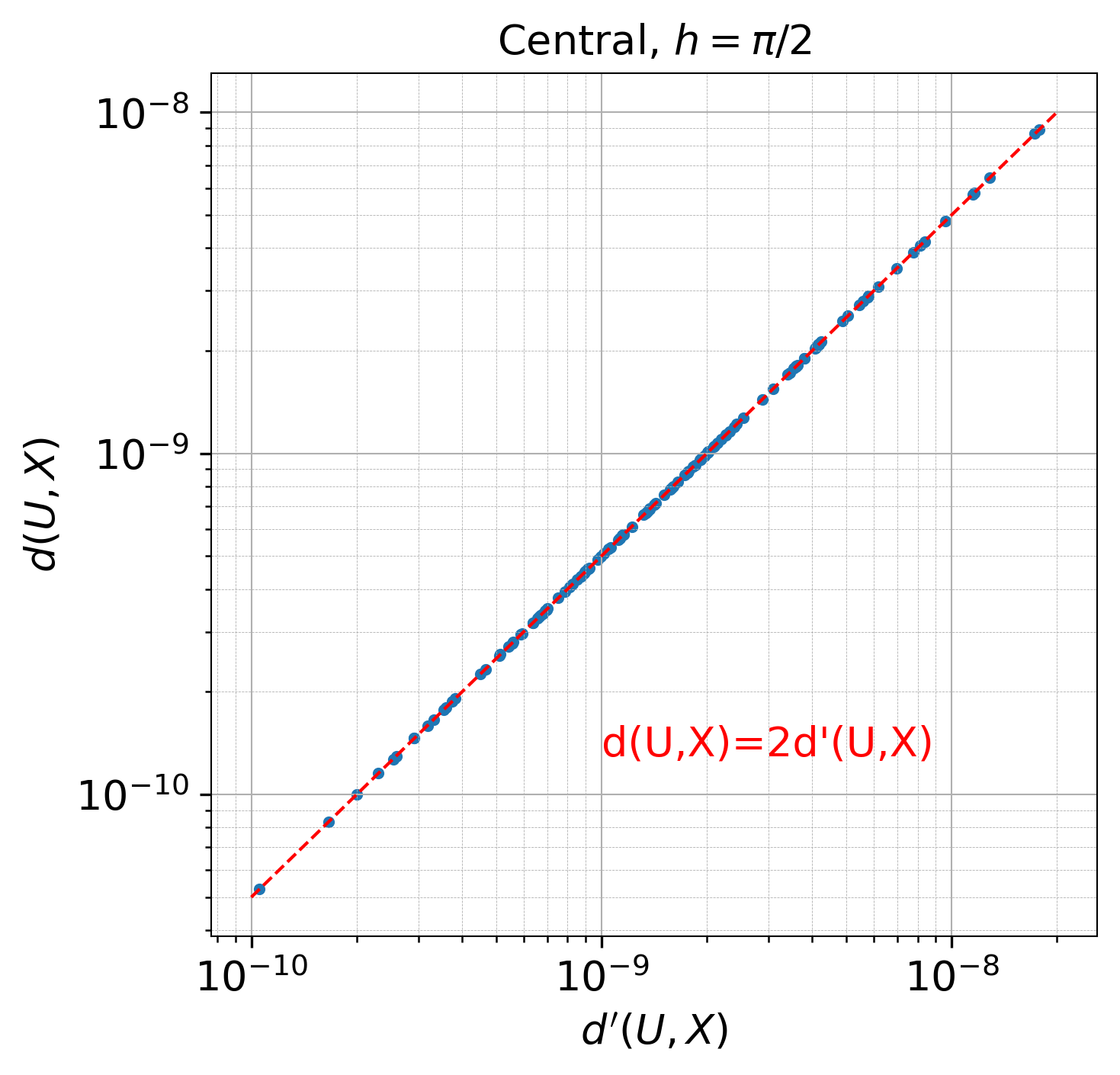}}
\caption{Comparison of final errors in distances after optimization. Each blue point represents one optimization trial, with a total of 128 trials plotted. The horizontal axis corresponds to $d'(U, X)$, and the vertical axis to $d(U, X)$}
\label{fig:phase_insensitive_relation}
\end{figure}

We analyze the convergence and final error of the proposed phase-insensitive distance $d'(U, X)$. Figure \ref{fig:phase_insensitive_convergence_comp} compares the convergence between intensity detection and coherent detection. The result for intensity detection shows good convergence, attributed to the unimodal property of $d'(U, X)$ and the optimization property of the few-layer redundant MPLC architecture \cite{Taguchi2023}. The optimization trace for intensity detection demonstrates faster convergence compared to coherent detection, which uses $\norm{U-X}_F^2$.
This faster convergence is attributed to the fewer number of phase shifters in the system using intensity detection, as depicted in Figure \ref{fig:MPLC}, which simplifies the optimization problem.
Figure \ref{fig:phase_insensitive_relation} compares final errors after optimization. Optimization is first performed with $d'(U, X)$, followed by calculating the value of $d(U, X)$ using the final optimized parameters.
The horizontal axis represents the proposed phase-insensitive distance $d'(U, X)$ defined in Eq. (\ref{eq:new_phase_insensitive}) and the vertical axis shows the previous distance $d(U, X)$ defined in Eq. (\ref{eq:prev_phase_insensitive}).
We conducted 128 optimizations, with each point in the figure corresponding to one optimization trial. A clear linear relationship $d'(U, X) = 2d(U, X)$ is observed.

\section{Conclusion}
In conclusion, we demonstrated that the exact gradient of the matrix norm for programmable unitary converters can be derived using the central difference method, without the need for any additional equipment. By leveraging a large finite difference, this technique effectively mitigates the influence of measurement noise. Furthermore, we introduced a phase-insensitive distance that yields the exact gradient when employing the proposed central difference method.
Thanks to the unimodal property of the phase-insensitive distance, few-layer redundant MPLC architecture can be configured using intensity detection.
These properties are rooted in the fact that a sweep of a single phase shifter in the system prompts a sinusoidal shift in the matrix norm's value.
This fact can be used as a reference to calibrate the phase shifters for the linearity.
We anticipate that this study will contribute to the robust and accurate unitary matrix realization in programmable unitary converters.

\begin{acknowledgments}
We wish to acknowledge Sho Yasui for the fruitful discussion. This work is supported by JST CREST Grant Number JPMJCR1872, JSPS KAKENHI JP23H00271, and JSPS KAKENHI JP22KJ1078.
\end{acknowledgments}

\appendix
\newpage
\section{Unimodality of $d'(U, X)$}
Here, we present a mathematical proof of the unimodality \cite{Taguchi2023} of the proposed norm $d'(U, X)$.

\textbf{Theorem} \textit{Given U, the function $v(X)=d'(U, X)$ has no local minimum on $U(N)$.}

\textbf{Proof.} Let $s_{ij} = \abs{\qty[XU^\dagger]_{ij}}$. The function $v(X)$ simplifies to:
\begin{equation}
\begin{aligned}
v(X) &= d'(U, X) = \sum_{ij} \qty| \delta_{ij} - s_{ij}^2 | \\
&=\begin{alignedat}{4}
& 1-s_{11}^2& &+s_{12}^2& &+\cdots& &+s_{1n}^2\\
& + s_{21}^2& &+1-s_{22}^2& &+\cdots& &+s_{2n}^2\\
& \vdots& &\vdots& &+1-s_{ii}^2& &+\cdots\\
& + s_{n1}^2& &+s_{n2}& &+\cdots& &+1-s_{nn}^2\\
\end{alignedat} \\[10pt]
&=N+\begin{alignedat}{4}
& s_{11}^2& &+s_{12}^2& &+\cdots& &+s_{1n}^2\\
+&s_{21}^2& &+s_{22}^2& &+\cdots& &+s_{2n}^2\\
& \vdots& &\vdots& &+s_{ii}^2& &+\cdots\\
+&s_{n1}^2& &+s_{n2}& &+\cdots& &+s_{nn}^2\\
\end{alignedat} - 2\sum_i s_{ii}^2 \\[10pt]
&=N+\norm{XU^\dag}^2_F-2\sum_i s_{ii}^2.
\end{aligned}
\end{equation}
Since both $X$ and $U^\dag$ are unitary matrices, we have $\norm{XU^\dag}^2_F = N$. Then, $v(X)$ is expressed as
\begin{equation}
\label{eq:v_simple}
v(X)=2N-2\sum_i s_{ii}^2.
\end{equation}
Finding the minimum of $v(X)$ is equivalent to finding the maximum of $\sum_i s_{ii}^2$. Without loss of generality, let $U=I$ (the identity matrix). To investigate the term $\sum_i s_{ii}^2$, we use the following theorem: for $n \geq 2$, a complex vector $(c_1, c_2, \cdots, c_n)$ is the diagonal of a unitary matrix if and only if the vector $\eta = (\abs{c_1}, \abs{c_2}, \cdots, \abs{c_n})$ satisfies the following conditions \cite{Horn1954,Tromborg1978}:
\begin{equation}
\label{eq:polyhedron}
\left\{
\begin{aligned}
&0 \leq \eta_i \leq 1 & (i = 1, \cdots, n), \\
&\sum_{i=1}^n \eta_i - 2\eta_l \leq n-2 & (l = 1, \cdots, n),
\end{aligned}
\right.
\end{equation}
where $\eta_i = \abs{c_i}$. This theorem indicates that the mapping of all unitary matrices through $\eta$ results in a convex polyhedron in $\mathbb{R}^n$, representing each unitary matrix as a point within this polyhedron \cite{Tromborg1978}. Let this polyhedron, defined in Eq. (\ref{eq:polyhedron}), be denoted as $\mathcal{P}$. Given $X$, the term $\sum_i s_{ii}^2$ is the square of the norm of a point in the convex polyhedron, that is, $\norm{p}^2$ for $p \in \mathcal{P}$. The maximum occurs at the point $p_\mathrm{max} = (1, 1, \cdots, 1) \in \mathcal{P}$. For any point $p \neq p_\mathrm{max} \in \mathcal{P}$, a dividing point $p_\mathrm{div}=(1-\varepsilon)p+\varepsilon p_\mathrm{max},\ 0 \leq \varepsilon < 1$ also lies within $\mathcal{P}$ due to its convexity. Since $\norm{p_\mathrm{div}}^2 < \norm{p_\mathrm{max}}^2$, $\norm{p}^2$ cannot be a maximum. Therefore, the square of the norm $\norm{p}^2 = \sum_i s_{ii}^2$ reaches its global maximum at $p = p_\mathrm{max}$, with no other local maxima in $\mathcal{P}$. Taking into account the continuity of the map $U \in U(N) \mapsto (\abs{U_{11}}, \abs{U_{22}}, \cdots, \abs{U_{nn}}) \in \mathbb{R}^n$, we now conclude that $v(X)$ has no local minimum. Q.E.D.

\bibliography{apssamp}%

\end{document}